# X-Rel: Energy-Efficient and Low-Overhead Approximate Reliability Framework for Error-Tolerant Applications Deployed in Critical Systems

Jafar Vafaei, Omid Akbari, Muhammad Shafique, Christian Hochberger

*Abstract*— Triple Modular Redundancy (TMR) is one of the most common techniques in fault-tolerant systems, in which the output is determined by a majority voter. However, the design diversity of replicated modules and/or soft errors that are more likely to happen in the nanoscale era may affect the majority voting scheme. Besides, the significant overheads of the TMR scheme may limit its usage in energy consumption and area-constrained critical systems. However, for most inherently error-resilient applications such as image processing and vision deployed in critical systems (like autonomous vehicles and robotics), achieving a given level of reliability has more priority than precise results. Therefore, these applications can benefit from the approximate computing paradigm to achieve higher energy efficiency and a lower area. This paper proposes an energy-efficient appro<u>x</u>imate <u>rel</u>iability (X-Rel) framework to overcome the aforementioned challenges of the TMR systems and get the full potential of approximate computing without sacrificing the desired reliability constraint and output quality. The X-Rel framework relies on relaxing the precision of the voter based on a systematical error bounding method that leverages user-defined quality and reliability constraints. Afterward, the size of the achieved voter is used to approximate the TMR modules such that the overall area and energy consumption are minimized. The effectiveness of employing the proposed X-Rel technique in a TMR structure, for different quality constraints as well as with various reliability bounds are evaluated in a 15-nm FinFET technology. The results of the X-Rel voter show delay, area, and energy consumption reductions of up to 86%, 87%, and 98%, respectively, when compared to those of the state-of-the-art approximate TMR voters. Also, the effectiveness of the proposed X-Rel-based TMR structure is assessed in four benchmark applications from different domains. For these benchmarks, results show 1.59×, 2.35×, and 3.39× Energy-Delay-Area-Product (EDAP) reduction for less than 1%, 5%, and 10% output quality degradations, respectively. Finally, an image processing application is benchmarked to evaluate the X-Rel framework efficacy in presence of errors, where the results show up to a 4.78× higher output image quality in comparison with the typical TMR voters.

J. Vafaei is with the School of Electrical and Computer Engineering, University of Tehran, Tehran 14395-515, Iran (e-mail: jafar.vafaei@ut.ac.ir).
O. Akbari (corresponding author) is with the Department of Electrical and Computer Engineering, Tarbiat Modares University, Tehran, Tehran 14115-111, Iran (e-mail: o.akbari@modares.ac.ir).
M. Shafique is with the Division of Engineering, New York University Abu Dhabi (NYU AD), Abu Dhabi 129188, United Arab Emirates (e-mail: muhammad.shafique@nyu.edu).
Ch. Hochberger is with the Department of Electrical Engineering, TU Darmstadt, 64289 Darmstadt, Germany (e-mail: hochberger@rs.tu-darmstadt.de).

"This work was supported in parts by the NYUAD's Research Enhancement Fund (REF) Award on "eDLAuto: An Automated Framework for Energy-Efficient Embedded Deep Learning in Autonomous Systems", and by the NYUAD Center for CyberSecurity (CCS), funded by Tamkeen under the NYUAD Research Institute Award G1104."

## I. Introduction

Soft errors are an important phenomenon in the nanoscale era that may affect the reliability of digital systems, in particular those used in critical applications [1]. This is worsened by technology scaling, as a greater level of integration occurs and the operating voltage of integrated circuits (ICs) is scaled while the capacitance of nodes is reduced. These factors decrease the charge threshold required to affect the device's normal operation and may corrupt the data [1]. Recently, researchers have proposed several techniques to mitigate the soft error effects (event upsets) at different hardware/software layers, which are mainly based on redundancy schemes [2]-[4].

N-Modular Redundancy (NMR) is a well-known technique for this purpose, in which the main module is employed along with *N-1* identical replicated modules [5]. This scheme needs at least *(N+1)/2* error-free modules to provide the correct output, i.e., up to *(N-1)/2* module failure is tolerable. However, to avoid common-mode failures (CMFs), design diversity can be utilized, where the replicated modules with different implementations are used. Thus, with the occurrence of a CMF, the replicated modules may result in different outputs and the error is therefore detected [6]. This is while, in a fault-free condition, the different implementations may produce slightly different outputs, which is considered an error by the TMR voter. Consequently, typical TMR voters cannot properly address the CMF problem [7]. Furthermore, the replication of modules significantly increases the design metrics, including delay, area, and energy consumption, while increasing the size and number of modules leads to a more complex voter with lower reliability [8],[29],[34].

An emerging approach for reducing the energy consumption of the computation and required area, as well as increasing the performance, is approximate computing, also known as inexact computing [9][10]. This type of computing provides the aforementioned advantages in error-resilient applications at the cost of inducing some errors. Approximate computing has been of great interest in recent years, e.g., being employed in on-chip artificial intelligence (AI) accelerators by IBM and attracting attention in designing low-power processors by ARM [9]. Digital signal processing (DSP), image and video processing, computer vision, and machine learning are examples of error-resilient applications [11]. These applications can tolerate a small accuracy relaxation at the output since a span of outputs is acceptable to still work functionally correct rather than a unique (golden) answer. As an example, a source of error resilience in an image processing application is that the output image is evaluated by the human eyes which are not very sensitive to small quality loss. Approximate computing leverages this accuracy relaxation in the computations of error-resilient applications as a knob for creating a trade-off between the quality and design metrics of the target application.

However, error-resilient applications exploited in critical systems may tolerate a given level of approximation with sustaining the reliability requirements, i.e., having a desired level of reliability has a higher priority than generating precise results [4],[26]-[28]. For instance, in autonomous vehicles, the necessary information must be extracted and analyzed from the images received from the surrounding environment to prevent lane departure. However, in this case, the reliable computation has a higher priority than the exact computation of unnecessary details [4]. Thus, an interesting research venue is to employ approximate computing in error-resilient applications deployed in critical systems [18]-[24],[26]-[28]. One notices that this method cannot be employed in those parts of a system that may undermine its functionality, e.g., in the control unit, where all of the bits are equally important.

This paper proposes an approximate reliability framework (X-Rel) to compose an energy-efficient approximate TMR-based design, including approximate voter and approximate modules, which meets both quality and reliability constraints. The design flow of X-Rel is based on a systematic approach that receives the voter minimum output quality by the user, and translates it to the amount of precision relaxing of the voter. This step results in a simpler voter with improved design metrics, including performance, area, and energy consumption, while retaining the reliability requirement, as well as resolving the challenges of the typical TMR voters. Then, the size of the designed voter is used to approximate the TMR modules to minimize the overall area and energy consumption. Approximate modules are achieved based on a proposed integer linear programming (ILP) formulation, which creates a tradeoff between the quality constraint and energy consumption. Our novel contributions in a nutshell are:

1) Proposing a systematic approach for designing an approximate TMR-based system, including the voter and modules, which operates based upon the user-defined quality and reliability constraints resulting in a low overhead TMR-based system.

2) Proposing an approximate TMR voter with a substantially lower delay, area, and energy consumption, compared to those of the typical and state-of-the-art approximate TMR voters such as the reduced precision redundancy (RPR)-based majority voters.

3) Creating a tradeoff between the reliability and implementation cost of the error-tolerant applications, by employing the ILP formulations for designing the modules of the TMR-based system.

4) Resolving the challenges of typical TMR voters, such as soft error effects, while retaining the reliability, as well as increasing the output quality in comparison with state-of-the-art approximate TMR voters.

5) Exploring the different design metrics such as delay, power, energy, and Energy-Delay-Area-Product (EDAP) of the X-Rel-based TMR designs for the various values of output quality degradations.

6) Analyzing the soft error tolerance and implementation cost tradeoff of the proposed X-Rel voter compared to the typical TMR voter.

The rest of this paper is organized as follows. Related works on approximate computing and reliable systems utilizing this computing paradigm are reviewed in Section II. The proposed X-Rel framework is presented in Section III. Section IV provides the design metrics evaluation of the X-Rel-based voter and compares it with the state-of-the-art approximate voters. Also, the effectiveness of the proposed framework is assessed via different benchmark applications. Moreover, the efficacy of the X-Rel framework against error is assessed in an image processing application in this section. Finally, the paper is concluded in Section V.

## II. RELATED WORK

In this section, first, we review the literature on approximate computing to find a general insight into approximate computing. Afterwards, we discuss those works that employ the approximate computing paradigm in reliable systems.

### A. Approximate Computing

Recently, several works have focused on employing approximate computing at different system layers, including hardware [10]-[13], architecture [14][15], and software layers [16][17]. A retrospective and prospective view of approximate computing can be found in [9].

In [10], an $n$-bit adder is segmented into $m$ sub-adders, where the input carry of each segment is speculated by its previous segment. Therefore, the carry chain is limited to two segments. A reconfigurable approximate carry look-ahead adder (RAP-CLA) was proposed in [11]. In this adder, a conventional CLA was transformed into two approximate and supplementary parts, which provide the ability to switch between exact and approximate operating modes. In [12], the structure of approximate logarithmic representation-based multipliers has been studied. Four approximate 4:2 compressors with the ability to operate in both exact and approximate modes have been proposed in [13]. A compressor is composed of two serially connected full adders for compressing four input bits into two output bits. The effectiveness of the proposed compressors in [13] was studied by using them in the structure of a Dadda multiplier.

At the architecture level, an approximate coarse-grained reconfigurable architecture (X-CGRA) has been proposed in [14], which provides the ability to implement an application with various quality levels to achieve different levels of energy saving. An approximation-aware ISA that supports dual-voltage operations to switch between exact and approximate operating modes was proposed in [15]. In [16], a code perforation technique has been proposed that reduces the computation overheads and saves power by skipping some iterations in a loop. In particular, this technique can be employed for error-resilient applications containing intensive loops, such as Monte Carlo simulations. A language-level approximation technique was introduced in [17], which uses type qualifiers to specify the non-critical data as the candidates of approximation, and then can separate the exact and approximate parts of a program.

A comprehensive survey on hardware approximate techniques for deep neural network accelerators can be found in [41]. In [42], a quality-aware approximation method for artificial neural networks (ANNs) has been proposed, which approximates the computation and memory accesses of certain less critical neurons while meeting a given quality constraint to achieve a quality-energy tradeoff. A layer-wise approximation method has been proposed in [43]. In this paper, the multipliers inside the neurons are approximated based on a genetic optimization procedure, and without retraining. In [44], a wight-oriented approximation method has been proposed for NNs to increase energy efficiency. In this method, approximate multipliers are employed to dynamically adjust the accuracy level of NNs during runtime, while meeting the accuracy loss threshold. This method also does not need intensive NN retraining.

*B. Approximate Computing in Fault-Tolerant Systems*

An extensive survey on approximate computing and its fault-tolerance property was presented in [18]. Also, exhaustive reviews on approximate computing techniques applied in fault-tolerant schemes to mitigate cost overheads has been conducted in [19][30]. An approximate TMR scheme has been proposed in [20], which is mainly based on successive approximation and loop perforation methods. Successive approximation is composed of loop-based algorithms whose output accuracy is increased in each iteration. Thus, by varying the number of iterations, a tradeoff between quality and performance emerges.

The challenges of selective hardening for arithmetic circuits to create a tradeoff between reliability and cost have been discussed in [21]. In [22], a Boolean factorization-based method has been presented to compose approximate modules of a TMR scheme. A transistor-level analysis for fault-tolerant voters was presented in [23], which is based on evaluating all possible states of the voter inputs. Also, this paper used a quadded transistor redundancy approach to mask the transient faults in the voter inputs. A partial TMR scheme for FPGAs was proposed in [24], which is mainly based on using approximate logic to compose the replicated modules. The flexibility of the proposed method and its fine granularity create an adaptive trade-off between reliability and hardware overheads.

In [25], a selective hardening TMR approach was proposed, which creates cost-effective redundant hardware. In [26], an approximate gate library was combined with a multi-objective optimization genetic algorithm to optimize the fault coverage of the TMR-based designs and lower their area overhead by changing the logic gates with the approximate ones. The proposed method in [27] employs approximate logic in the fault mitigation area, specifically for FPGAs, where their flexibility could be employed to create an optimal trade-off between overheads and reliability. The work of [28] also tries to create a tradeoff between the reliability and area overheads of TMR-based designs. However, to generate an optimal redundant logic based on the approximate logic, this work compared two evolutionary and probabilistic approaches. A method to lower the overheads of TMR-based designs is employing the RPR scheme, in which a full precision module is used along with the two reduced precision modules. However, one of the major issues with the RPR scheme is its complex voter and consequently imposed overheads, which is due to the required extra hardware for comparing the results of full precision and reduced precision modules with each other, and also with a predetermined threshold for generating the final output [8],[29]. In [35], a circuit-level power gating-based RPR scheme for unsigned integer arithmetic units was proposed. In this method, the full precision module is powered off, and in case the difference between the outputs of the reduced precision modules is higher than a threshold, the full precision module is turned on to achieve higher accuracy. In [8], a voting-based RPR structure was proposed for adders. This method splits the adder structure (main module) into two MSB and LSB sub-adders, where the redundant modules in the TMR structure are composed of only MSB adders resulting in area and power improvements. Also, by forwarding the output carry of the main module LSB adder to the redundant MSB adders, the accuracy is improved. However, this method is applicable only to the carry propagate adders (CPAs), in particular to the ripple carry adders (RCAs), where there is no accuracy analysis or solution for bounding the quality loss.

Almost all of the previously studied works are only applicable to a given TMR-based design, e.g., the work of [8] is appropriate for the RCAs. However, to the best of our knowledge, the works of [6] and [7] are state-of-the-art works that have proposed a generic approximate voter independent of the implemented modules by the N-modular redundancy (NMR) scheme. Therefore, to have a fair comparison we will employ these works in our comparisons with state-of-the-art approximate voters (see the last column of TABLE II) in Section IV, wherein the following, we study their structure in detail.

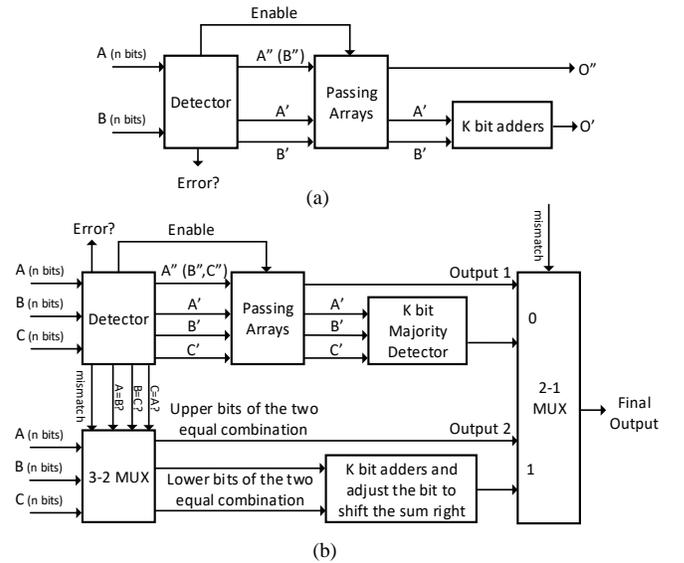

Fig. 1. State-of-the-art inexact voters, a) Inexact Double Modular Redundancy (IDMR) Voter of [6], b) Inexact Triple Double Modular Redundancy (ITDMR) Voter of [7].

An inexact double modular redundancy (IDMR) was proposed in [6] (see Fig. 1.a). IDMR splits the $n$-bit inputs ($A$ and $B$) of a voter into two parts, including lower $k$ bits ($A'$ and $B'$) and upper $n-k$ bits ($A''$ and $B''$). Next, in the *detector* block, it calculates the difference between the voter inputs. If the calculated difference is smaller than a predefined threshold, the inputs are considered valid and the *Enable* signal is set to one. Otherwise, when the difference is larger than the threshold, an *Error* signal is generated. The *Passing Arrays* block which is composed of an array of AND gates, propagates the inputs when the *Enable* signal is activated. Then, the upper $n-k$ bits of one of the voter inputs and the average of inputs lower $k$ bits (calculated by the *k-bit adders* block) are used as the voter output upper and lower bits, respectively. However, employing these components in the IDMR voter results in a more complex structure rather than a typical TMR voter, and consequently, excessive delay, area, and energy consumption overheads [8]. In specific, the typical TMR voter may be faster than the IDMR voter since it does not require any adders [6].

The IDMR approach was extended to the TMR voters in [7], which was entitled inexact TMR-DMR (ITDMR). Similar to the IDMR, ITDMR compares the three inputs of the voter based on their differences and produces the output. In cases all pairwise differences between upper $n-k$ bits of the inputs are larger than the threshold, an *Error* signal is generated (see Fig. 1.b). Otherwise, several extra components such as a complex *3-2 multiplexer*, *k-bit adders*, and a *2-1 multiplexer* are employed to determine the output (see Fig. 1.b). Note that the employed *3-2 multiplexer* in the IDTMR structure is composed of several subtractors and comparators to select the proper result (see [7] for more details).

Unlike state-of-the-art techniques that either target only approximating voter (such as [6][7]) or only approximating certain redundant modules (like [8][35]), in this paper, we aim at targeting full system approximations where both redundant compute modules and voter can be approximated to leverage the full potential of approximate computing. However, this comes with certain additional challenges, e.g., meeting the user-defined output quality constraint and removing the required complex voters in the RPR-based designs. To the best of our knowledge, this paper is the first comprehensive framework that approximates the whole TMR resolving the aforementioned challenges.

## III. PROPOSED X-REL FRAMEWORK

In this section, the X-Rel framework, including the design steps of the approximate voter and modules is discussed. For this, we first discuss details of an exact TMR voter. Next, the structure and details of the proposed voter are explained.

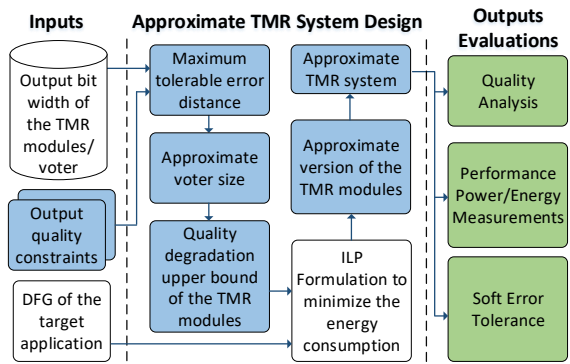

Fig. 2. System diagram of the proposed XRel framework from inputs to outputs evaluations.

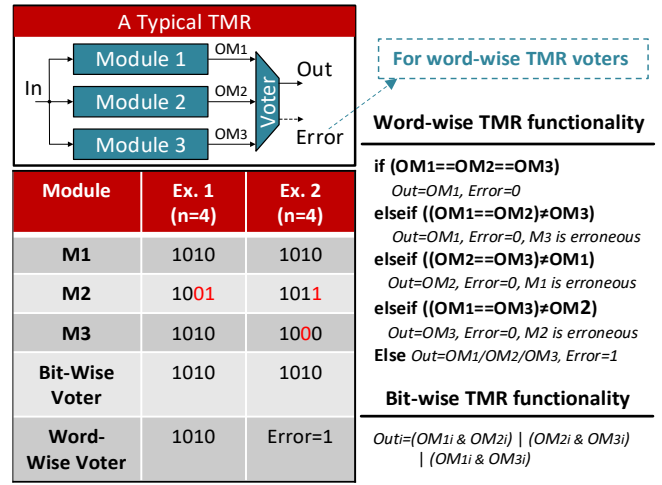

Fig. 3. A typical TMR structure in two bit-wise and word-wise schemes, along with examples to show their difference.

Afterwards, the method of designing the approximate modules considering the obtained voter of the previous steps is introduced. Fig. 2 shows the system diagram of the proposed X-Rel framework including, inputs, approximate TMR system design, and outputs evaluations. Different segments of this framework are discussed in the following.

### A. Exact Majority-Based TMR Voter

A typical TMR scheme is composed of three replicated modules along with a majority-based voter (see Fig. 3), in which at least two modules have to operate correctly to result in a valid output by the voter. The TMR scheme may be applied to the entire system or selectively to the critical parts. Also, two bit-wise and word-wise schemes can be considered for the TMR voter to compose the output. Examples provided in Fig. 3 show the differences between these schemes. However, a bit-wise TMR voter is not efficient in terms of data integrity [7]. Also, word-wise TMR is sensitive to a slight difference between the voter inputs, which is referred to as *strict majority* property [6]. Fig. 4 shows an example of a simple word wise-voter employed for majority voting of a module with 2-bit output ($S$, $C$), along with generating *Error* signal when none of the voter input pairs are equal [40].

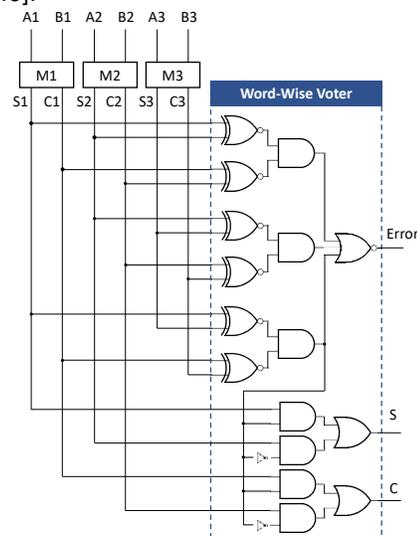

Fig. 4. An example word-wise voter employed for TMR majority voting of a module with 2-bit output [40].

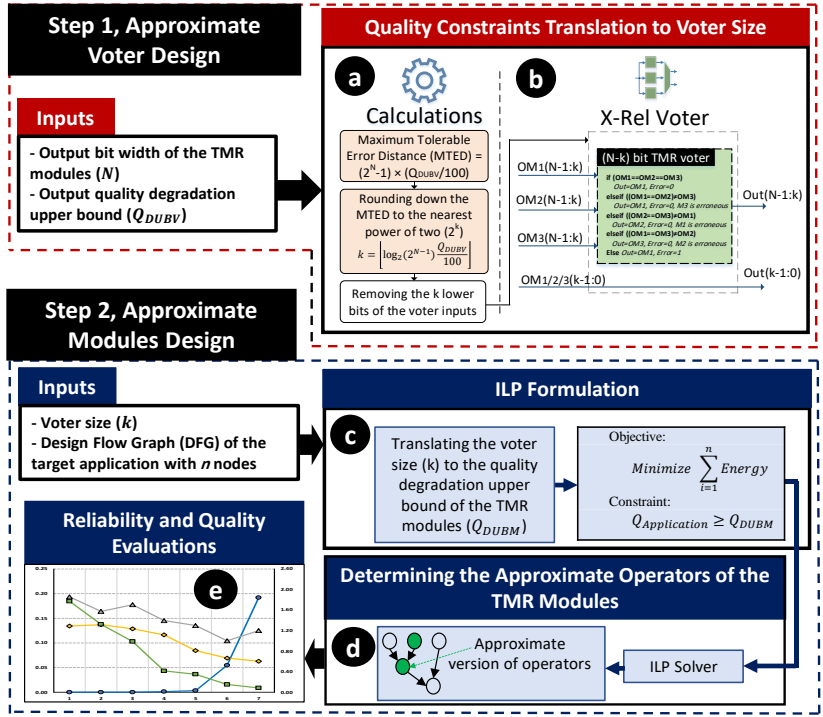

Fig. 5. The proposed X-Rel framework, including the design steps of the approximate voter and modules.

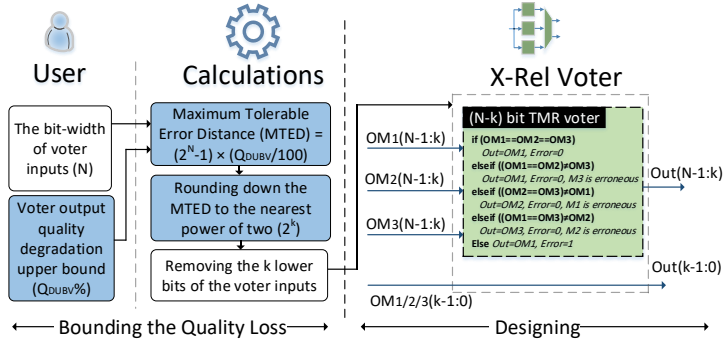

Fig. 6. X-Rel voter design flow, along with an example to show its design method and functionality.

*B. X-Rel Approximate voter*

Fig. 5 illustrates the proposed X-Rel design steps, reaching from the approximate voter to the approximate modules. As shown in this figure, the proposed design flow for constructing the X-Rel voter is based on receiving the quality constraint and the voter inputs size from the user, and then, translating them to the size of a voter which is smaller than the primary one (typical voter). The main objective of this translation is to meet the user-defined quality constraint by the X-Rel voter. However, for translating this constraint to the amount of precision relaxing at the voter inputs, we need an error metric that measures the amount of error at the voter outputs. There are different error metrics to evaluate the accuracy of approximate computations [11], e.g., error rate (ER), error distance (ED), mean relative error distance (MRED), and error variance ($v$). Formulations and descriptions of these metrics are shown in TABLE I. Each metric has its own purpose, e.g., in [11], ER was used to state the accuracy of arithmetic adders. We use the ED metric to design the X-Rel voter since it operates based on the differences between exact ($O$) and approximate ($O'$) values, which means that the ED is independent of the implemented applications by the TMR scheme. The X-Rel voter design flow is shown on the left side of Fig. 6, which is composed of bounding the quality loss and designing steps. The inputs of this flow are the voter inputs bit-width ($N$) and voter output quality degradation upper bound ($Q_{DUBV}\%$) constraint, where the user is responsible for defining these inputs. Next, the inputs are employed to calculate the size of the X-Rel voter (see bounding the quality loss segment in Fig. 6). To this aim, at first, the $Q_{DUBV}$ is used to calculate the proposed maximum tolerable error distance (MTED) at the voter output which is defined by

$$MTED = (2^N - 1) \times \frac{Q_{DUBV}}{100} \quad (1)$$

Note that the user can also directly define the MTED for a given application. Then, the obtained MTED from (1) is rounded down to the nearest power of two, which is in a $2^k$ format. This rounding down guarantees that the ED of the X-Rel voter never violates MTED, and therefore, the $Q_{DUBV}$ constraint is always satisfied. Thus, the parameter $k$ can be obtained by

$$k = \lfloor \log_2 MTED \rfloor \quad (2)$$

Now, the obtained value of $k$ corresponds to the number of lower bits in the voter inputs that do not affect the $Q_{DUBV}$ (can even be considered as don't care bits). Hence, $k$ lower bits of voter inputs are relaxed (see the designing step in Fig. 6). To improve the accuracy, different schemes can be employed to compose the $k$ lower bits of the proposed approximate voter, e.g., using a $k$-bit majority detector such as the one employed in [7] (see Fig. 1.b), or setting the $k$ lower bits to a constant value (zero or one). In the X-Rel voter structure, the $k$ lower bits of one of the voter inputs are forwarded to the output directly, which may lead to a higher accuracy without the need for extra hardware and its related overheads (e.g., delay, power, and area). Although the selected method may lead to a large error distance in the $k$ lower bits, since the X-Rel voter design flow is based on the MTED (see (1)), the induced error is still in the tolerable range. Moreover, we assumed that each of the three replicated modules of an X-Rel-based design is affected by the noise (getting faulty) with the same probability. Therefore, without loss of generality, it is up to the designer to select the lower part of which module to be forwarded to the output. Furthermore, in some cases, e.g., the ripple-carry adders (RCAs) [8], the designer may decide to use the lower bits of one of the voter inputs and simplify the structure of two other modules in a TMR-based design in such a way that they only generate upper bits.

TABLE I  SOME ERROR METRICS FOR QUALITY EVALUATION OF APPROXIMATE SYSTEMS

| Error Metric | Formulation | Description |
|---|---|---|
| ER | $\dfrac{\#erroneous\ outputs}{T}$ | $T$: Total number of output samples |
| ED | $\|O - O'\|$ | $O$: Exact Output $O'$: Aprx. output |
| MRED | $\dfrac{1}{T}\sum_{i=1}^{T}\dfrac{\|O_i - O'_i\|}{O_i}$ | - |
| $v$ | $\dfrac{1}{T-1}\sum_{i=1}^{T}(O_i - O'_i)^2$ | - |

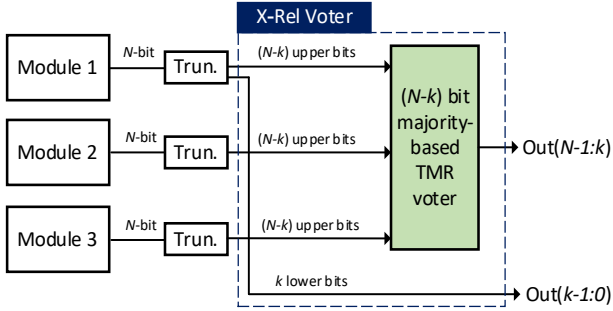

Fig. 7. The X-Rel voter with the size of $N - k$.

As shown in Fig. 7, using the $k$ lower bits of one of the voter inputs leads to an X-Rel voter with the size of $N - k$, which depending on the value of $k$ may result in more efficient design metrics than a typical voter with the size of $N$. An example of the proposed design flow is shown on the right side of Fig. 6, where for $Q_{DUBV} = 10\%$ (up to 10% quality loss is allowed at the output) and $N = 8$, the proposed design flow results in $k = 4$. Therefore, an approximate voter with a size of 4, i.e. $(N - k)$ can be used instead of a typical voter with a size 8, i.e. $N$. Also, to illuminate the practicality of the proposed X-Rel voter, a numerical example is depicted in Fig. 6, where some bits of the voter inputs are considered erroneous (red-colored bits). The results show that the X-Rel voter provides an approximate output with $ED = 2$, which meets the user-defined $Q_{DUBV}$ given in (2). Moreover, a single error in the $N - k$ upper bits of the inputs ($OM_3$) has been tolerated, which shows that the reliability of the TMR structures is retained by the X-Rel voter. Note that, for the example shown in Fig. 6, a typical TMR voter with the *strict majority* property cannot find an output. Moreover, in case there is no error in LSB bits, X-Rel voter produces an output with the same quality as the typical TMR voter.

Finally, as discussed in Section II, the state-of-the-art approximate TMR voters use extra hardware such as subtractors and comparators to compute the output [6][7]. However, this makes their voter more complex than a typical TMR voter and imposes excessive delay, area, and energy consumption overheads. Instead, as shown in this subsection, we proposed a systematic error bounding method for designing the X-Rel voter as a major advantage to benefit the error-resiliency of critical applications and mitigate the *strict majority* problem.

### C. X-Rel Approximate Modules

In this subsection, based on the designed X-Rel voter, the modules of the TMR-based design are approximated such that the energy consumption of the overall design is minimized compared to a conventional TMR-based design.

As shown in step 2 of Fig. 5, inputs of the approximate modules design step are the X-Rel voter size $(N - k)$ and the Data Flow Graph ($DFG(V, E)$) of the target application, where $V$ and $E$ are the node set and their data dependency, respectively. First, based on the X-Rel voter bit width, the number of the output bits of the TMR modules that can be relaxed (quality of the voter inputs) is determined, which is equal to $k$. As an example, for a module with 32-bit output, and for $k = 4$, four LSB bits of each approximated module of the TMR-based design can be relaxed, i.e., the maximum ED of the approximate modules is $2^4$ ($2^k$). Now, we have to determine which nodes of the DFG should be approximated, such that the module final output quality meet the obtained maximum ED. The key issue for this step is to assess the impact of approximating each DFG node at the module output. To tackle this problem, and minimize the energy consumption of the TMR-based design, we define a constrained optimization problem as follows.

Objective:
$$Minimize \sum_{i=1}^{n} E_i \qquad (3)$$

Constraint:
$$Q_O \geq 1 - Q_{DUBM}$$

where $E_i$ is the energy consumption of the $i^{th}$ node of the DFG, and $n$ is the total number of nodes. Also, $Q_O$ and $Q_{DUBM}$ are the output quality and quality degradation upper bound of a single module, respectively, which are numbers between 0 and 1. To solve the optimization problem defined (3), different methods are conceivable, such as evolutionary algorithms like the genetic algorithm (GA), Integer Linear Programming (ILP) formulation, and greedy algorithms like the Knapsack problem. However, evolutionary algorithms and, more generally, nature-inspired metaheuristics like GA and swarm optimization do not guarantee to find of the optimal solution due to their probabilistic nature of the solution [45]. Other optimization algorithms like Knapsack which is resolved by a greedy method, also cannot guarantee the optimal solution for the 0/1 problem [46]. Whereas in the case of linear programming, a global optimum will always be attained [45]. Thus, we selected the ILP formulation to attain the optimal solution. For this target, the constraint of (3) can be described based on the $v$ (error variance), which is a function of squared ED (see TABLE I), and provides a proper error propagating scheme to assess the computational error of approximate nodes at the module output [14][36]. The output variance ($v_o$) of a DFG is obtained by [36]

$$v_o = \sum_{i=1}^{n} ES_i^2 \cdot v_i \qquad (4)$$

where $ES_i$ and $v_i$ are the error sensitivity and error variance of the $i^{th}$ node, respectively. Moreover, $ES_i$ is calculated as [36]

$$ES_i = \frac{\epsilon_{i,o}}{\epsilon_i} \qquad (5)$$

where $\epsilon_i$ and $\epsilon_{i,o}$ are the error of approximating the $i^{th}$ node, at the $i^{th}$ node and DFG output, respectively. Note that $\epsilon_i$ and $\epsilon_{i,o}$ are the mean error distances calculated by injecting a set of uniform random inputs to the application. Also, based on the approximation technique employed for the DFG nodes, error variance can be calculated. In this work, we employ the truncation technique as a widely used approximate computing method to achieve area and energy consumption saving for the addition and multiplication units in the target DFG [37]. In the truncation technique, an approximate ($a$-$j_a$)-bit adder is implemented instead of an exact $a$-bit adder when its $j_a$ lower bits are truncated. Also, an approximate ($m$-$j_m$)-bit multiplier is implemented instead of an exact m-bit multiplier, when its $j_m$ LSB bits are truncated. These lower sizes of approximate components may result in considerably lower delay, area, and power/energy for the X-Rel-based design rather than the typical TMR structure.

As mentioned before, the maximum ED of the approximate module is $2^k$. Thus, error variance upper bound ($v_{UB}$) can be written as

$$v_{UB} = \frac{1}{N-1} \sum_{i=1}^{N}(O_i - O_i')^2 = \frac{1}{N-1} \sum_{i=1}^{N}(2^k - 1)^2$$
$$= \frac{N}{N-1}(2^k - 1)^2 \qquad (6)$$

Now, by considering the error variance as the quality metric of approximate modules in the constraint of (3), and by using (4) and (6), we have

$$\sum_{i=1}^{n} ES_i^2 \cdot v_i \leq \frac{N}{N-1}(2^k - 1)^2 \qquad (7)$$

By using (7), the ILP formulation of (3) is written as

Objective:

$$Minimize \sum_{j=0}^{l}\sum_{i=1}^{n} E_{(i,j)} \cdot x_{(i,j)} \qquad (8)$$

Constraint:

$$\sum_{j=1}^{l}\sum_{i=1}^{n} ES_{(i,j)}^2 \cdot v_{(i,j)} \cdot x_{(i,j)} \leq \frac{N}{N-1}(2^k - 1)^2$$

where, $E_{(i,j)}$ is the energy consumption of the $i^{th}$ node when its $j$

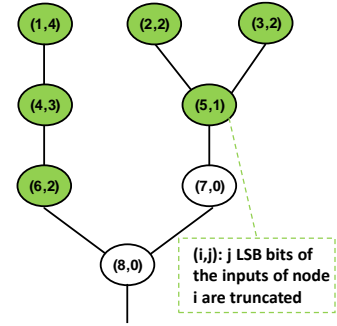

Fig. 8. An example of a sample DFG when the proposed X-Rel approximate module design step is applied on.

lower bits are truncated, and $l$ is the bitwidth of this node. $x_{(i,j)}$ is a pseudo-Boolean variable to determine the operating mode of the $i^{th}$ node, i.e., the $i^{th}$ node will operate under a $j$ LSB bits truncation. As an example, when $x_{(i,2)}$ is one (zero) the $i^{th}$ node is implemented such that its 2 lower bits are truncated. Also, to ensure that the $i^{th}$ node is implemented in the exact mode or is truncated only for a given number of LSB bits, the following constraint is defined:

$$\sum_{j=0}^{l} x_{(i,j)} = 1 \qquad (9)$$

where $x_{(i,0)}$ means that the $i^{th}$ node is implemented without truncation, i.e., the exact mode. It is worth noting that for a DFG with multiple outputs, the proposed accuracy constraints should be defined for each output, separately. Finally, by employing the proposed ILP formulations, a DFG of an application can be implemented in the different accuracy modes, as well as the various area and energy consumption levels. Fig. 8 shows an example of applying the proposed X-Rel approximate module design step on a sample DFG, in which by solving the proposed ILP formulas in (8)-(9), different numbers of the LSB bits of green-colored nodes were truncated. The obtained approximate version of a given application can be used for implementing the modules of the X-Rel-based TMR design. Note that by using the proposed ILP formulation, some LSB bits of the arithmetic units inside the TMR modules may be approximated (truncated). In this case, the accuracy of the $k$ lower bits may be lower than the typical exact TMR module. However, the value of $k$ is obtained based on $Q_{DUBV}$ (see (1) and (2)). Thus, as mentioned in Subsection III.B, the user must define the $Q_{DUBV}$ such that the approximate computations still produce results within the boundary of the sufficient required quality.

One may use finer granularity level redundancy techniques for each truncated/untruncated arithmetic unit of the target DFG. Such a hardening technique can be applied to all the arithmetic units, or selectively to only some of them. Moreover, employing the redundancy techniques at the finer granularity level, without triplication the whole design as a TMR structure is possible. In this case, the proposed ILP method is still applicable for implementing the application. First, by using the proposed ILP formulation in (8), some of the arithmetic units of the target DFG may be truncated, while the use-defined quality constraint is met. Then, the considered redundancy technique at the finer granularity level is applied to the approximated/exact arithmetic units of the design.

TABLE II KEY ATTRIBUTES OF THE DIFFERENT STUDIED STATE-OF-THE-ART APPROXIMATE NMR-BASED DESIGNS

| Ref. | Proposed Method | Cons | Proposing a method to design inexact voter | Selected to Compare |
|---|---|---|---|---|
| [6] | An inexact double modular redundancy voter | High design metrics (area, delay, power) Without error masking capability | ✓ | ✓ |
| [7] | An inexact triple double modular redundancy voter | High design metrics | ✓ | ✓ |
| [8] | TMR based Carry Propagate Adder (CPA) | Applicable only to the carry propagate adders | × | × |
| [20] | Loop perforation in loop-based algorithms | Applicable only to loop-based algorithms | × | × |
| [21] | Using approximate logics for composing redundant modules | Applicable only to the arithmetic units | × | × |
| [22] | Boolean factorization-based method to compose approximate redundant modules | Applicable only to the arithmetic units | × | × |
| [23] | Using pass transistors and quadded transistor-level redundancy to achieve a fault-tolerant voter | Transistor-level voter design High design metrics due to the quadded transistor-level redundancy | ✓ | × |
| [24] | Composing partial TMR circuits for FPGAs using approximate logic circuits | Appropriate for FPGA-based designs | × | × |
| [25] | Employing a selective hardening redundancy in a circuit instead of hardening an entire circuit | - | × | × |
| [26] | Combining the approximate logics with the optimization genetic algorithm | - | × | × |
| [27] | Composing partial TMR circuits for FPGAs using approximate logic circuits based on testability analysis | Appropriate for FPGA-based designs | × | × |
| [28] | Evaluating the evolutionary and probabilistic approaches for using approximate logics in the TMR-based designs | - | × | × |
| [35] | A circuit-level power gating-based RPR scheme to power off the full precision module in case the difference of the reduced precision modules is higher than a given threshold. | - | × | × |

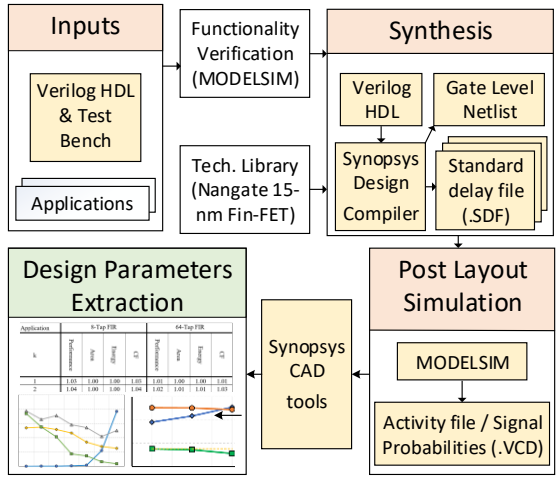

Fig. 9. The developed tool flow and simulation setup for the synthesis of the different studied designs.

## IV. RESULTS AND DISCUSSION

In this section, first, we compare the design metrics of the proposed X-Rel voter with state-of-the-art approximate voters. Then, the efficacy of the proposed X-Rel framework is evaluated for different benchmark applications in various application domains. Afterwards, we investigate the effectiveness of the proposed approximate X-Rel framework against error in an image processing application. Then, we analyze the soft error tolerance and implementation cost tradeoff of the proposed X-Rel voter compared to the typical TMR voter. All the designs have been implemented in Verilog HDL and synthesized using Synopsys Design Compiler with a 15nm FinFET-based Open Cell Library (OCL) technology at the operating voltage of 0.8V [31]. In the syntheses, the *compile_ultra* command was used to ungroup all components and automatically synthesize circuits based on the timing constraints. Fig. 9 shows the tool-flow and simulation setup employed to extract the design metrics of the different designs in this section.

### A. Design metrics of the X-Rel-based Voter

TABLE II shows the key features of the different studied NMR-based designs, which were studied in Section II. As shown in this table, the works proposed in [6] (i.e., IDMR), [7] (i.e., ITDMR), and [23] are only state-of-the-art works that proposed a method to design approximate voters independent of the implemented modules by the NMR scheme. Also, as discussed in Section II, the proposed method in [23] is applicable at the transistor level and has focused on using pass transistors and quadded transistor-level redundancy to achieve a fault-tolerant voter. Therefore, our proposed X-Rel voter and the work of [23] are complementary and not competitive. As an example, the proposed method of X-Rel voter can be used to design a voter that meets the user-defined quality constraint, whereas, without loss of generality the method of [23] can be employed to mask the internal faults of the X-Rel voter. Therefore, two state-of-the-art voters of the IDMR and ITDMR, which proposed an approximate voter independent of the implemented modules, along with the typical TMR voter were employed in our comparisons.

The results for the delay, power, area, and energy-delay product (EDP)×area of the investigated voters for different values of $k$ and $N$ are shown in Fig. 10. Note that in our simulations, we used a typical $(N-k)$-bit majority-based TMR voter in the structure of the proposed X-Rel (the green box in Fig. 7), in which at least two modules have to operate correctly for generating the correct output (see Subsection III.A). The results indicate that, for various values of $k$, compared to the typical 8-bit (16- and 32-bit) TMR voter, the X-Rel improves, on average, the delay 12% (6% and 6%), power 21% (27% and 16%), area 25% (23% and 25%), energy 29% (32% and 21%), and EDP×area 44% (46% and 39%), respectively. Since the proposed X-Rel voter is designed by relaxing the voter inputs precision (see Subsection III.B), all of its design metrics have been improved compared to those of the typical TMR voter. However, in comparison with the typical TMR voter, for the various studied values of $k$ and $N$, the ITDMR (IDMR) voter worsens, on average, the delay 389% (403%), power 419% (127%), area 371% (101%), energy 2,535% (1135%), and EDP×area 12,938% (2477%), respectively.

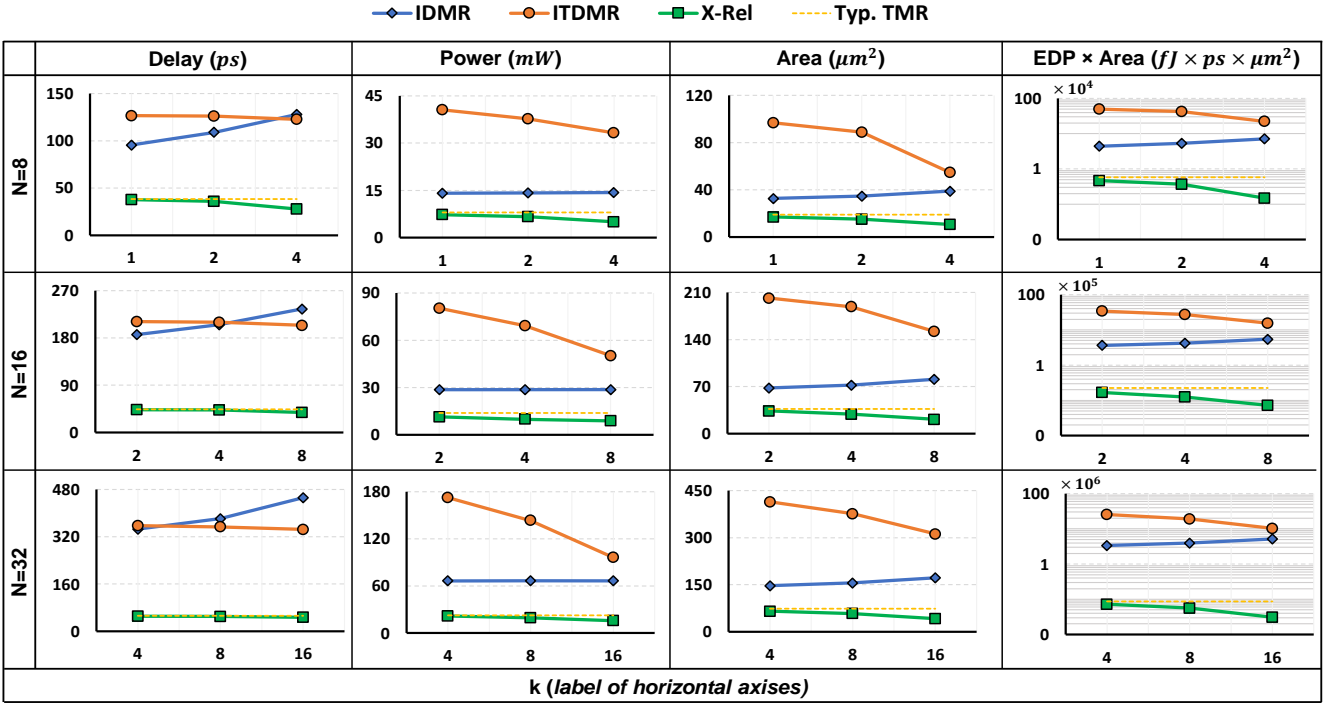

Fig. 10. Design metrics of the Typical TMR, IDMR, ITDMR, and X-Rel Voters for Different values of $k$ and $N$.

It should be noted that most of the IDMR design metrics are increased with the increment of $k$. The reason is that the IDMR has a high dependency on the adder utilized in its architecture, where proportional to the $k$'s increment, the size of the adder is enlarged. However, with the increase of $k$, ITDMR design metrics are reduced. The reason is that the ITDMR voter has employed some complex components such as a detector to determine the pairwise difference between its inputs, which need several subtractors and comparators (see Subsection II.B). In general, as also pointed out in [8], the required error detection and steering logic make most of the state-of-the-art voter schemes inefficient in design metrics. In contrast, our proposed X-Rel voter overcomes this challenge efficiently.

### B. Evaluation of X-Rel-based TMR Designs

To investigate the effectiveness of the proposed X-Rel-based TMR designs, four benchmarks from different application domains are examined, including 8- and 64-Tap finite impulse response (FIR) filters [14], 8 × 8 Matrix Multiplication (MM), and Smoothing filter (with a 3 × 3 filter) (SMT) [38]. As discussed in Section III.C, for the different applications, by varying the parameter $k$ the different number of operations may be approximated. Specifically, the ILP formulations proposed in (7)-(9) can result in the approximated operations while keeping the output quality degradation under the desired threshold. In this work, the LINGO software [39] was used to solve the aforementioned ILP problems. Note that the inputs of the LINGO software were generated by a MATLAB-based in-house developed tool.

In the performed evaluations, $N$ was considered as 16. Also, the $Q_{DUBV}$ was in the range of 0.006% to 12.5%. TABLE III shows the obtained values for the parameter $k$ and $v_{UB}$ calculated using (2) and (6), respectively. TABLE IV shows the design metrics improvements of the investigated applications under the various values of the parameter $k$ obtained from TABLE III. The design metrics are performance (delay), area, energy, and energy-delay-area product (EDAP), which were obtained for the parameter $k$ ranging from 1 to 12. Note that the improvement ratios are obtained by comparing the design metrics of the obtained approximate versions of each studied application with the those of exact one (typical TMR). Moreover, the works of [6] and [7] have nothing to do with the modules of a TMR structure, and thus, were not considered for comparisons with the results of TABLE IV.

TABLE III OBTAINED VALUE OF THE PARAMETER K AND $v_{UB}$ FOR THE DIFFERENT VALUES OF $Q_{DUB}$ AND $N = 16$

| $Q_{DUBV}$ (%) less than | $k$ | $v_{UB}$ |
|---|---|---|
| 0.006 | 1 | 1.06E+00 |
| 0.012 | 2 | 9.60E+00 |
| 0.024 | 3 | 5.22E+01 |
| 0.048 | 4 | 2.40E+02 |
| 0.097 | 5 | 1.03E+03 |
| 0.195 | 6 | 4.23E+03 |
| 0.390 | 7 | 1.72E+04 |
| 0.781 | 8 | 6.94E+04 |
| 1.562 | 9 | 2.79E+05 |
| 3.125 | 10 | 1.17E+06 |
| 6.250 | 11 | 4.47E+06 |
| 12.500 | 12 | 1.79E+07 |

TABLE IV PERFORMANCE, AREA, ENERGY, AND EDAP IMPROVEMENTS OF THE STUDIED APPLICATION BENCHMARKS IMPLEMENTED BASED ON THE X-REL FRAMEWORK UNDER THE VARIOUS VALUES OF THE PARAMETER $k$ FROM 1 TO 12 WHEN COMPARED TO THE $k=0$ (TYPICAL TMR)

| Application | Improvements (×) | | | | | | | | | | | | | | | |
|---|---|---|---|---|---|---|---|---|---|---|---|---|---|---|---|---|
| | 8-Tap FIR | | | | 64-Tap FIR | | | | 8 × 8 MM | | | | SMT | | | |
| k | Perform. | Area | Energy | EDAP | Perform. | Area | Energy | EDAP | Perform. | Area | Energy | EDAP | Perform. | Area | Energy | EDAP |
| 1 | 1.03 | 1.00 | 1.00 | 1.03 | 1.01 | 1.00 | 1.00 | 1.01 | 1.00 | 1.00 | 1.00 | 1.00 | 1.00 | 1.00 | 1.00 | 1.00 |
| 2 | 1.04 | 1.00 | 1.00 | 1.04 | 1.02 | 1.01 | 1.01 | 1.03 | 1.03 | 1.00 | 1.00 | 1.03 | 1.03 | 1.00 | 1.00 | 1.03 |
| 3 | 1.06 | 1.00 | 1.00 | 1.07 | 1.02 | 1.01 | 1.02 | 1.06 | 1.05 | 1.00 | 1.00 | 1.06 | 1.05 | 1.00 | 1.00 | 1.06 |
| 4 | 1.08 | 1.00 | 1.01 | 1.09 | 1.03 | 1.03 | 1.04 | 1.11 | 1.07 | 1.00 | 1.01 | 1.08 | 1.07 | 1.00 | 1.01 | 1.08 |
| 5 | 1.10 | 1.01 | 1.01 | 1.12 | 1.05 | 1.05 | 1.08 | 1.19 | 1.10 | 1.01 | 1.01 | 1.11 | 1.10 | 1.01 | 1.01 | 1.12 |
| 6 | 1.13 | 1.01 | 1.01 | 1.15 | 1.04 | 1.09 | 1.13 | 1.27 | 1.13 | 1.01 | 1.01 | 1.15 | 1.15 | 1.01 | 1.01 | 1.17 |
| 7 | 1.17 | 1.05 | 1.07 | 1.30 | 1.09 | 1.12 | 1.18 | 1.44 | 1.16 | 1.01 | 1.01 | 1.19 | 1.17 | 1.01 | 1.01 | 1.19 |
| 8 | 1.19 | 1.09 | 1.13 | 1.46 | 1.14 | 1.15 | 1.21 | 1.59 | 1.19 | 1.05 | 1.07 | 1.33 | 1.17 | 1.04 | 1.06 | 1.29 |
| 9 | 1.23 | 1.13 | 1.21 | 1.69 | 1.13 | 1.20 | 1.29 | 1.75 | 1.21 | 1.18 | 1.29 | 1.85 | 1.15 | 1.18 | 1.29 | 1.74 |
| 10 | 1.28 | 1.18 | 1.28 | 1.94 | 1.13 | 1.35 | 1.53 | 2.35 | 1.26 | 1.28 | 1.44 | 2.32 | 1.23 | 1.27 | 1.42 | 2.22 |
| 11 | 1.33 | 1.23 | 1.35 | 2.21 | 1.17 | 1.57 | 1.85 | 3.39 | 1.28 | 1.48 | 1.77 | 3.37 | 1.30 | 1.45 | 1.71 | 3.23 |
| 12 | **1.39** | 1.23 | 1.36 | 2.32 | 1.23 | **2.07** | **2.64** | **6.72** | 1.36 | 1.74 | 2.25 | 5.32 | 1.32 | 1.72 | 2.23 | 5.09 |

Based on the results, increasing $Q_{DUBV}$ ($k$) improves the performance (delay), area, energy, and EDAP of the studied X-Rel-based TMR applications. Specifically, for the investigated applications, there are delay, area, energy, and EDAP reductions of up to 1.39×, 2.07×, 2.64×, and 6.72×, respectively, all in the case of up to 12.5% of $Q_{DUBV}$ ($k = 12$). Also, for the four different studied applications and the $Q_{DUBV} = 12.5\%$, the delay, area, energy, and EDAP metrics are reduced, on average, by 1.32×, 1.69×, 2.12×, and 4.86×, respectively. Note that based on the results provided in TABLE IV, a higher k results in more improvements in the studied metrics. However, as discussed in Subsection III.B, the maximum allowable value of $k$ is determined based on the user-defined $Q_{DUBV}$.

Finally, to evaluate the effectiveness of the proposed ILP formulation in (6)-(9), we calculated the output variances of the four investigated applications under the different values of $v_{UB}$ presented in TABLE III. The results of this study have been shown in TABLE V, where almost all the quality constraints were satisfied. However, only in one case, for the 64-Tap FIR application and when $v_{UB}$ is 2.40E+2 ($k = 4$), the output variance is 2.5% higher than the variance upper bound. This violation is due to the intrinsic error of (4) [36][14]. Note that the ILP is an NP-Complete problem, where the required time to find the optimal solution is increased with the growing number of variables and constraints [14]. For the examined benchmarks, the number of constraint equations (application nodes) is in the range of 14 to 127. Based on the simulations, the used ILP solver (Lingo [39]) found the solutions in less than 12.3 seconds, where the worst case belongs to the 64-tap FIR filter application benchmark. Thus, heuristic approaches can be used when the runtime becomes too much.

TABLE V OUTPUT VARIANCE OF THE STUDIED APPLICATIONS UNDER THE DIFFERENT VARIANCE UPPER BOUNDS

| $v_{UB}$ | $v_o$ | | | |
|---|---|---|---|---|
| | 8-Tap FIR | 64-Tap FIR | 8 × 8 MM | SMT |
| 1.06E+00 | 4.78E−01 | 9.39E−01 | 8.84E−01 | 9.44E−01 |
| 9.60E+00 | 4.43E+00 | 8.92E+00 | 6.05E+00 | 7.14E+00 |
| 5.22E+01 | 4.79E+01 | 5.02E+01 | 4.79E+01 | 4.89E+01 |
| 2.40E+02 | 2.29E+02 | **2.46E+02** | 2.04E+02 | 2.18E+02 |
| 1.03E+03 | 9.84E+02 | 9.91E+02 | 8.59E+02 | 9.09E+02 |
| 4.23E+03 | 4.07E+03 | 4.10E+03 | 4.04E+03 | 3.37E+03 |
| 1.72E+04 | 1.63E+04 | 1.66E+04 | 1.38E+04 | 1.45E+04 |
| 6.94E+04 | 6.58E+04 | 6.71E+04 | 6.42E+04 | 6.71E+04 |
| 2.79E+05 | 2.62E+05 | 2.69E+05 | 2.63E+05 | 2.69E+05 |
| 1.17E+06 | 1.06E+06 | 1.08E+06 | 1.06E+06 | 1.08E+06 |
| 4.47E+06 | 4.30E+06 | 4.32E+06 | 4.32E+06 | 4.21E+06 |
| 1.79E+07 | 1.45E+07 | 1.73E+07 | 1.73E+07 | 1.72E+07 |

### C. FPGA Implementations

In this subsection, we evaluate and compare different configurations of our proposed X-Rel voter, the studied state-of-the-art approximate voters, and the typical TMR voter for performance and area (#LUTs). To this aim, we synthesized the Verilog HDL of the studied voters in XILINX Vivado, for Kintex-7 XC7K70T-2FBV676 FPGA. Also, the clock frequency has been set to 125 MHz employing the *Explore* strategy provided by the Vivado tool. TABLE VI shows the results of this exploration. Based on the results, for the different values of *k* and *N*, the X-Rel voter provides, on average, 57% (54%) lower delay compared to the IDMR (ITDMR) voter. In terms of area, the X-Rel voter results in, on average, 42% (73%) lower number of LUTs compared to the IDMR (ITDMR) voter. Also, the X-Rel voter offers a 3% (26%) improvement in delay (area) rather than the typical TMR voter.

Moreover, we synthesized the studied applications in Subsection IV.B using XILINX Vivado for Kintex-7 XC7K70T-2FBV676 FPGA. TABLE VII shows the results of this exploration for a wide range of k for performance, area, and delay-area-product (DAP) compared to the typical TMR structure. Based on the results, for the investigated applications, X-Rel achieved delay, area, energy, and DAP reductions of up to 1.11×, 2.07×, and 2.21×, respectively. Also, for $Q_{DUBV}$=12.5%, the delay, area, energy, and EDAP metrics are reduced, on average, by 1.08×, 1.61×, and 1.75×, respectively.

### D. Image Processing Application

To study the efficacy of the proposed X-Rel voter, we use it in an image processing application. To evaluate its performance against errors, we need a controllable error insertion model, which introduces some errors at the voter inputs. For this purpose, we use the noise model shown in Fig. 11 inspired by

TABLE VI  DESIGN METRICS OF THE TYPICAL TMR, IDMR, ITDMR, AND X-REL VOTERS FOR DIFFERENT VALUES OF $k$ AND N, WHEN IMPLEMENTED ON XILINX KINTEX-7 XC7K70T-2FBV676 FPGA

| Metric | N=8 | | | | | | N=16 | | | | | | N=32 | | | | | |
|---|---|---|---|---|---|---|---|---|---|---|---|---|---|---|---|---|---|---|
| | Delay (ns) | | | Area (#LUT) | | | Delay (ns) | | | Area (#LUT) | | | Delay (ns) | | | Area (#LUT) | | |
| k | 1 | 2 | 4 | 1 | 2 | 4 | 2 | 4 | 8 | 2 | 4 | 8 | 4 | 8 | 16 | 4 | 8 | 16 |
| TMR | | 1.49 | | | 18 | | | 1.93 | | | 34 | | | 2.03 | | | 66 | |
| IDMR | 2.65 | 2.61 | 3.77 | 14 | 15 | 35 | 3.10 | 3.90 | 4.91 | 61 | 50 | 74 | 5.42 | 5.91 | 7.86 | 76 | 104 | 145 |
| ITDMR | 3.11 | 2.98 | 2.84 | 45 | 41 | 49 | 3.60 | 3.80 | 3.94 | 96 | 94 | 93 | 4.71 | 4.82 | 5.34 | 196 | 228 | 195 |
| X-TMR | 1.57 | 1.30 | 1.40 | 17 | 14 | 11 | 2.01 | 1.86 | 1.70 | 30 | 25 | 18 | 2.02 | 2.06 | 1.98 | 59 | 49 | 34 |

TABLE VII  PERFORMANCE, AREA, AND DAP IMPROVEMENTS OF THE X-REL FRAMEWORK FOR STUDIED APPLICATION BENCHMARKS IMPLEMENTED ON XILINX KINTEX-7 XC7K70T-2FBV676 FPGA, UNDER THE VARIOUS VALUES OF K COMPARED TO THE TYPICAL TMR

| Application | 8-Tap FIR | | | 64-Tap FIR | | | 8×8 MM | | | SMT | | |
|---|---|---|---|---|---|---|---|---|---|---|---|---|
| k | Performance | Area (#LUT) | DAP | Performance | Area (#LUT) | DAP | Performance | Area (#LUT) | DAP | Performance | Area (#LUT) | DAP |
| 1 | 1.05 | 1.00 | 1.06 | 1.02 | 1.00 | 1.02 | 1.01 | 1.00 | 1.01 | 1.01 | 1.00 | 1.01 |
| 2 | 1.04 | 1.01 | 1.05 | 1.03 | 1.01 | 1.03 | 1.05 | 1.00 | 1.06 | 1.05 | 1.01 | 1.06 |
| 3 | 1.03 | 1.01 | 1.04 | 1.03 | 1.01 | 1.05 | 1.03 | 1.01 | 1.04 | 1.03 | 1.01 | 1.04 |
| 4 | 1.04 | 1.01 | 1.05 | 1.06 | 1.03 | 1.09 | 1.03 | 1.01 | 1.05 | 1.03 | 1.01 | 1.05 |
| 5 | 1.05 | 1.02 | 1.07 | 1.05 | 1.05 | 1.11 | 1.05 | 1.02 | 1.06 | 1.05 | 1.02 | 1.06 |
| 6 | 1.05 | 1.02 | 1.07 | 1.07 | 1.09 | 1.16 | 1.05 | 1.02 | 1.07 | 1.05 | 1.02 | 1.08 |
| 7 | 1.05 | 1.05 | 1.10 | 1.05 | 1.12 | 1.17 | 1.04 | 1.03 | 1.07 | 1.04 | 1.03 | 1.07 |
| 8 | 1.04 | 1.09 | 1.14 | 1.08 | 1.15 | 1.24 | 1.04 | 1.06 | 1.11 | 1.04 | 1.05 | 1.10 |
| 9 | 1.06 | 1.13 | 1.20 | 1.07 | 1.20 | 1.28 | 1.05 | 1.17 | 1.23 | 1.06 | 1.15 | 1.22 |
| 10 | 1.08 | 1.17 | 1.27 | 1.07 | 1.35 | 1.45 | 1.10 | 1.24 | 1.37 | 1.07 | 1.23 | 1.31 |
| 11 | 1.08 | 1.22 | 1.31 | 1.07 | 1.57 | 1.67 | 1.09 | 1.39 | 1.51 | 1.11 | 1.37 | 1.52 |
| 12 | 1.07 | 1.23 | 1.32 | 1.06 | **2.07** | **2.21** | 1.10 | 1.59 | 1.74 | **1.11** | 1.57 | 1.74 |

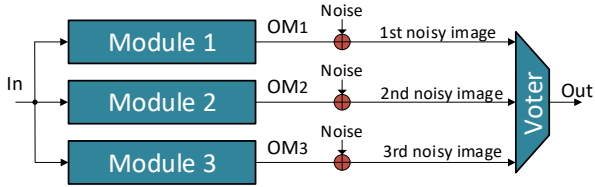

Fig. 11. The proposed noise model in [7] to insert errors at the input images of the voter.

[6][7], in which the noise sources are placed before the voter inputs. In this case, the injected errors are considered equivalent to the error occurrence inside of each of the replicated modules in the TMR-based design. Now, by controlling the amount of the inserted noise, various levels of erroneous voter inputs are achievable including the no error (exact) mode. One notices that the noise sources operate in a bit-wise manner that flips the bits of the modules' output (modules 1, 2, and 3) independently by the same flip probability ($P_f$). More precisely, corresponding to each bit of the module's outputs, a uniform random number is generated between 0 to 1. Afterward, if the generated number is less or equal to the considered $P_f$, then the corresponding bit is flipped. As an example, for a 24-bit pixel of an image and $P_f = 5\%$, 24 random values are generated between 0 to 1 and assigned to each bit of that pixel. Now, those bits whose corresponding random value is less than 0.05 are flipped.

Finally, to evaluate the quality of voter output images using the aforementioned noise model, we use the Mean Structural Similarity Index Metric (MSSIM). The MSSIM is calculated based on measuring the structural similarity of the exact and approximate images, which provides a better consistency with image quality perceived by humans, compared to the Peak Signal to Noise Ratio (PSNR) metric [32]. MSSIM is obtained by [32]:

$$MSSIM(X,Y) = \frac{(2\mu_x\mu_y + C_1)(2\sigma_{xy} + C_2)}{(\mu_x^2 + \mu_y^2 + C_1)(\sigma_x^2 + \sigma_y^2 + C_2)} \quad (10)$$

where $\mu_x$, $\mu_y$, $\sigma_x$, $\sigma_y$, and $\sigma_{xy}$ are the local means, standard deviations, and cross-covariance of images $X$ and $Y$. The MSSIM is a number between 0 to 1, where the higher values show higher quality. Note that the used noise model shown in Fig. 11 is independent of the image processing application, i.e., any image processing application such as Sharpening and Smoothing may be implemented by the TMR system. In this evaluation, ten 24 bits/pixel standard benchmark images of [33] were used as the output of the TMR modules. Afterwards, the input images of the voters were generated by the noise model introduced in Fig. 11. The average MSSIM of the output images of the different investigated voters, for ten repetitions of the simulations, are given in TABLE VIII. Note that the repetition of the simulations is due to get more general and representative results. Also, Fig. 12 shows the input and obtained output images of the investigated voters for $k = 4$ and two different values of $P_f$ (1% and 5%), when the "Female" benchmark image was used. Note that for the IDMR voter, only the first and second noisy input images were used.

TABLE VIII  AVERAGE MSSIM OF OUTPUT IMAGES FOR TEN REPETITIONS, UNDER THE DIFFERENT INVESTIGATED VOTERS, FOR TWO VALUES OF $p_f$

| **Benchmark Image** | $P_f$ | **Typ. TMR** | **IDMR** | **ITDMR** | **X-Rel** |
|---|---|---|---|---|---|
| Female (NTSC test image) | 1% | 0.87 | 0.74 | 0.86 | 0.96 |
| | 5% | 0.39 | 0.44 | 0.64 | 0.56 |
| Couple (NTSC test image) | 1% | 0.93 | 0.87 | 0.94 | 0.95 |
| | 5% | 0.51 | 0.56 | 0.77 | 0.49 |
| Female (from Bell Labs) | 1% | 0.51 | 0.27 | 0.41 | 0.93 |
| | 5% | 0.08 | 0.10 | 0.17 | 0.40 |
| Female | 1% | 0.79 | 0.60 | 0.75 | 0.97 |
| | 5% | 0.26 | 0.31 | 0.49 | 0.67 |
| House | 1% | 0.83 | 0.64 | 0.81 | 0.99 |
| | 5% | 0.24 | 0.29 | 0.50 | 0.77 |
| Tree | 1% | 0.79 | 0.63 | 0.77 | 0.98 |
| | 5% | 0.30 | 0.35 | 0.52 | 0.72 |
| Jelly beans | 1% | 0.66 | 0.42 | 0.60 | 0.97 |
| | 5% | 0.16 | 0.20 | 0.31 | 0.64 |
| Mandrill | 1% | 0.88 | 0.75 | 0.88 | 0.99 |
| | 5% | 0.38 | 0.44 | 0.64 | 0.83 |
| Sailboat on lake | 1% | 0.81 | 0.66 | 0.80 | 0.98 |
| | 5% | 0.34 | 0.39 | 0.56 | 0.73 |
| Peppers | 1% | 0.92 | 0.83 | 0.92 | 0.99 |
| | 5% | 0.50 | 0.56 | 0.74 | 0.88 |
| **Average** | **1%** | **0.80** | **0.64** | **0.77** | **0.97** |
| | **5%** | **0.32** | **0.36** | **0.54** | **0.67** |

| | Input Image | | | | | | | |
|---|---|---|---|---|---|---|---|---|
| | Error-free image ($P_f = 0$) | First Noisy Input Image | Second Noisy Input Image | Third Noisy Input Image | Error-free image ($P_f = 0$) | First Noisy Input Image | Second Noisy Input Image | Third Noisy Input Image |
| Input Image | | | | | | | | |
| | Typical TMR | IDMR | ITDMR | X-Rel | Typical TMR | IDMR | ITDMR | X-Rel |
| Output Image | | | | | | | | |
| MSSIM | 0.79 | 0.60 | 0.75 | 0.97 | 0.27 | 0.31 | 0.49 | 0.67 |

Fig. 12. Error-free image along with the output images of the different investigated voters, for the two values of $P_f$.

Based on the results, for the investigated benchmark images, compared to the ITDMR (IDMR), X-Rel results in, on average, 26% (51%) and 24% (82%) higher MSSIM for $P_f = 1\%$ and $P_f = 5\%$, respectively. Note that, we directly used the $k$ lower bits of one of the voter inputs ($OM_1$) for the $k$ lower bits of the voter output, whereas ITDMR has a different structure introduced in Subsection II.B. Therefore, the output images of the X-Rel voter outperform the ITDMR and IDMR voters. For the investigated benchmark images, X-Rel achieves, on average, 21% and 108% higher MSSIM for $P_f = 1\%$ and $P_f = 5\%$, respectively, when compared to the typical TMR. These improvements are due to the fact that the typical TMR has the *strict majority* voting property and generates an error signal for slight differences in its inputs. On the other hand, X-Rel can tolerate an error distance up to the MTED between the inputs. For the differences higher than the MTED, similar to the other investigated voters, X-Rel generates an error signal and sets the output to zero.

Furthermore, when two inputs of the X-Rel voter are affected by the noise such that their $n - k$ upper bits get the same erroneous value, then the error is not detected, which means that X-Rel tolerates up to one single error in the inputs. This property is similar to that of typical TMR voters, which cannot tolerate multiple errors. Hence, in these cases, the majority voter produces an erroneous value as the correct output (false positive). In our simulations, for $P_f = 0.01$ ($P_f = 0.05$), X-Rel generated 79 (1649) false positive among the 196,608 (number of pixels in the "Female" image: 256×256×3) voter outputs.

### E. FIR Filter

In this subsection, we investigate the efficacy of the proposed X-Rel framework for an 8-Tap FIR filter, in which the input data and the filter coefficients, and the output were quantized in terms of 32-bit. Inspired by [7], the input signal of FIR filters was randomly generated in the range of [-0.5, 0.5]. Moreover, the bit-wise noise injection method shown in Fig. 11 was used for this study. TABLE IX shows the obtained results. Based on the results, for the $P_f = 1\%$ ($P_f = 5\%$) the X-Rel voter results in higher PSNR, on average, 12% (7%), 12% (6%), and 5% (7%) compared to the TMR, ITDMR, and IDMR voters, respectively. Overall, by increasing the $k$, the PSNR increases for the X-Rel, IDMR, and IDTMR voters since the probability of generating the correct results is increased.

TABLE IX   AVERAGE PSNR OF STUDIED VOTERS FOR THE 8-TAP FIR FILTER UNDER THE BIT-WISE NOISE INJECTION, FOR THE TWO VALUES OF $P_f = 1\%$ AND $P_f = 5\%$

| | $P_f = 1\%$ | | | | $P_f = 5\%$ | | | |
|---|---|---|---|---|---|---|---|---|
| | k=1 | k=2 | k=4 | k=8 | k=1 | k=2 | k=4 | k=8 |
| **TMR** | 62.4 | | | | 54.5 | | | |
| **X-Rel** | 70.1 | 70.2 | 70.2 | 70.3 | 58.2 | 58.2 | 58.2 | 58.4 |
| **ITDMR** | 62.5 | 62.5 | 62.6 | 62.8 | 54.6 | 54.7 | 54.7 | 54.8 |
| **IDMR** | 66.8 | 66.8 | 66.9 | 67.3 | 54.4 | 54.4 | 54.5 | 54.7 |

### F. Soft Error Tolerance and Implementation Cost Tradeoff

In this subsection, we evaluate the impact of soft errors on the proposed X-Rel voter, where the noise model shown in Fig. 11 was employed to insert uniformly distributed random errors at the voter inputs. Also, the Mean Square Error (MSE) ratio was employed as the soft error tolerance metric [29], defined by

$$MSE\ Ratio = \frac{MSE_{X-Rel}}{MSE_{Typ}} \quad (11)$$

where $MSE_{X-Rel}$ and $MSE_{Typ}$ are the MSE of the X-Rel voter and typical TMR voter, respectively. One may notice that the lower MSE Ratio indicates the higher soft error tolerance of the proposed X-Rel voter compared to the typical TMR voter.

The analysis is this subsection is an efficient method to create a trade-off between the reliability and implementation cost of the proposed X-Rel when compared to a typical TMR voter. Fig. 13 shows the average MSE ratio of an 8-bit X-Rel voter normalized to the typical TMR voter, for a hundred repetitions under the various values of k ranging from 1 to 7, and three different values of $P_f$, including 0.01, 0.05, and 0.1. Also, the EDAP metric is employed as the hardware (implementation) cost parameter, which is shown in the secondary axis of Fig. 13. As shown in this figure, the MSE ratio is in the range of ~ 0.01 to 0.14, i.e., the proposed X-Rel voter can decrease the MSE by about 86% to 99%, compared to the typical TMR voter. Also, by increasing the $P_f$ from 0.01 (blue line with circles markers) to 0.1 (yellow line with diamond markers), the MSE-ratio is decreased, i.e., X-Rel voter shows a better performance in the environment with lower noise. However, still, the MSE of the proposed X-Rel voter for the noisier circumstances is still significantly higher than the typical voter (at least 86%). One notices that depending on the amount of noise in the environment that the implemented application may be utilized, which can be modeled by the $P_f$, different tradeoffs are achievable by the proposed approximate voter. Therefore, the lower (higher) $P_f$ represents the quiet (noisy) circumstances in that the system may be employed in them, and the desired level of reliability is achievable.

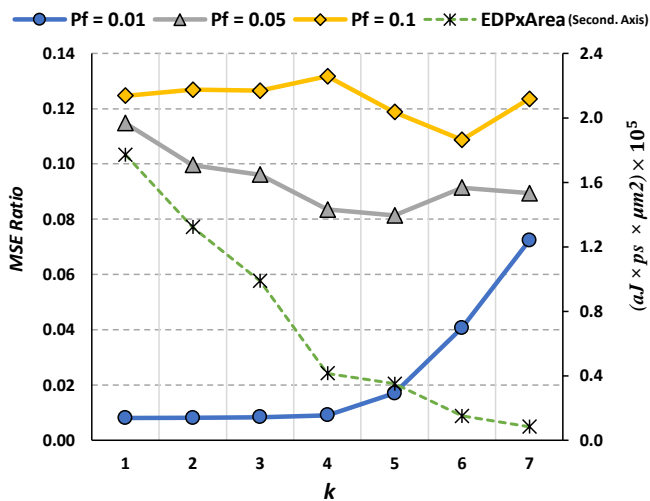

Fig. 13. Average MRED ratio of the X-Rel voter to the typical TMR voters, for 100 repetitions, under $N=8$ and the different values of $k$ and $P_f$.

## V. CONCLUSION

We presented an approximate reliability framework to mitigate the challenges of typical TMR-based designs and reduce their complexity depending on the tolerable error at the voter output. First, the proposed X-Rel voter is designed by using a systematic error bounding approach, in which the voter output quality degradation upper bound is received from the user and translated to the number of lower bits that can be relaxed in the voter inputs. Using this method, X-Rel removes the excessive overheads of the state-of-the-art approximate voters. Afterwards, the bit width of the achieved approximate voter is used for composing approximate modules of the TMR-based design such that the total energy consumption of the design is minimized. Also, the output quality of each replicated module remains higher than the one required to keep the voter output quality at the desired level. Based on the simulation results, compared to the typical TMR voter, X-Rel-based voters achieved up to 27%, 44%, and 54% improvements in the delay, area, and energy consumption, respectively. Also, the TMR-based structures composed based on the proposed framework achieved, on average, 1.7× lower EDAP for a user-defined quality degradation upper bound in the range from 0.006% to 12.5%, for the several studied benchmarks when compared to the exact TMR-based designs. Finally, the efficacy of the X-Rel framework against error was investigated by deploying the proposed voter in the image processing application. The results showed up to 114% (26%) MSSIM improvement in the X-Rel output images, compared to the typical TMR (state-of-the-art ITDMR) voter. In general, the X-Rel framework can be used as an appealing redundancy scheme for error-resilient critical applications.

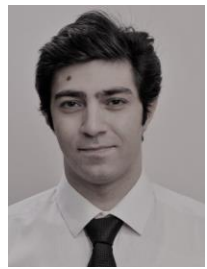

**Jafar Vafaei** received the B.Sc. degree from the University of Tabriz, Tabriz, Iran, in 2013, and the M.Sc. degree from University of Tehran, Tehran, Iran, in 2019, both in Electrical Engineering. He is currently a research assistant at Tarbiat Modares University, Tehran, Iran in the Computer Architecture and Dependable systems Laboratory (CADS-Lab). His research interests include low power digital designs and machine learning, reconfigurable computing, neuromorphic computing, and fault-tolerant system design.

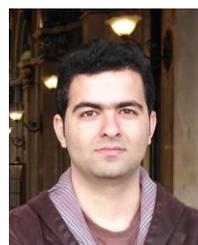

**Omid Akbari** received the B.Sc. degree from the University of Guilan, Rasht, Iran, in 2011, the M.Sc. degree from Iran University of Science and Technology, Tehran, Iran, in 2013, and the Ph.D. degree from the University of Tehran, Iran, in 2018, all in Electrical Engineering, Electronics - Digital Systems sub-discipline. He was a visiting researcher in the CARE-Tech Lab. at Vienna University of Technology (TU Wien), Austria, from Apr. to Oct. 2017, and a visiting research fellow under the Future Talent Guest Stay program at Technische Universität Darmstadt (TU Darmstadt), Germany, from Jul. to Sep. 2022. He is currently an assistant professor of Electrical and Computer Engineering at Tarbiat Modares University, Tehran, Iran, where he is also the Director of the Computer Architecture and Dependable Systems Laboratory (CADS-Lab). His current research interests include embedded machine learning, reconfigurable computing, energy-efficient computing, distributed learning, and fault-tolerant system design.

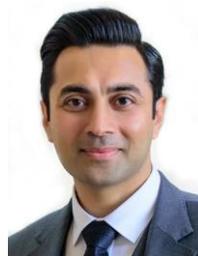

**Muhammad Shafique** (SM'16) received the Ph.D. degree in computer science from the Karlsruhe Institute of Technology (KIT), Germany, in 2011. In Oct.2016, he joined the Institute of Computer Engineering at the Faculty of Informatics, Technische Universität Wien (TU Wien), Vienna, Austria as a Full Professor of Computer Architecture and Robust, Energy-Efficient Technologies. Since Sep.2020, Dr. Shafique is with the New York University (NYU), where he is currently a Full Professor and the director of eBrain Lab at the NYU-Abu Dhabi in UAE, and a Global Network Professor at the Tandon School of Engineering, NYU-New York City in USA. He is also a Co-PI/Investigator in multiple NYUAD Centers. His research interests are in AI & machine learning hardware and system-level design, brain-inspired computing, quantum machine learning, cognitive autonomous systems, wearable healthcare, energy-efficient systems, robust computing, hardware security, emerging technologies, FPGAs, MPSoCs, and embedded systems.

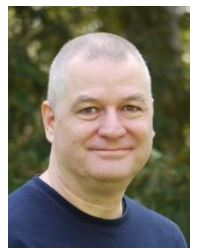

**Christian Hochberger** holds the chair for Computer Systems in EE department at TU Darmstadt since 2012. Before, he was an associate professor for embedded systems in CS department at TU Dresden since 2003. He got his diploma and PhD in computer science in 1992 and 1998. His research is focused on (re)configurable technology and design automation for such technology.